\documentclass[reprint,prl]{revtex4}
\usepackage[T1]{fontenc}
\usepackage[latin9]{inputenc}
\usepackage{color}
\usepackage{amsmath}
\usepackage{amssymb}
\usepackage{graphicx}
\usepackage{esint}

\makeatletter
\@ifundefined{textcolor}{}
{
 \definecolor{BLACK}{gray}{0}
 \definecolor{WHITE}{gray}{1}
 \definecolor{RED}{rgb}{1,0,0}
 \definecolor{GREEN}{rgb}{0,1,0}
 \definecolor{BLUE}{rgb}{0,0,1}
 \definecolor{CYAN}{cmyk}{1,0,0,0}
 \definecolor{MAGENTA}{cmyk}{0,1,0,0}
 \definecolor{YELLOW}{cmyk}{0,0,1,0}
 }

\begin{document}

\title{Comment on ``Fringe Visibility and Which-Way Information: An Inequality''}
\author{F. De Zela}
\affiliation{Departamento de Ciencias, Secci\'{o}n F\'{i}sica,
Pontificia Universidad Cat\'{o}lica del Perú, Ap. 1761, Lima,
Peru.}

\maketitle

An increasing number of recent papers that address Bohr's
complementarity refer to the so-called wave-particle duality
relation between ``distinguishability'' $\mathcal{D}$ and
visibility $\mathcal{V}$:
\begin{equation}
\mathcal{D}^2+\mathcal{V}^2\leq 1.\label{1}
\end{equation}
Relation (\ref{1}) was first derived by Jaeger, Shimony and
Vaidman \cite{Jaeger}. It was also derived, independently from the
latter, by Englert, who employed a more straightforward approach
\cite{Englert}. This last author, besides stressing the physical
meaning of (\ref{1}), also stressed that it is logically
independent of the uncertainty relation. It is thus important to
make sure that the reported derivations of relation (\ref{1}) are
free from any logical or technical flaw. Unfortunately, this is
not the case with Englert's derivation, which contains a slight
technical flaw. The purpose of this Comment is to point out this
flaw and to repair it. Neither the physical content of (\ref{1})
nor its validity is thereby questioned.

The proof of (\ref{1}), as presented in \cite{Englert}, is based
on the following inequality:
\begin{equation}
\mathcal{D}^2+\mathcal{V}^2\leq \sum_{j,k}D_{j} D_{k}
\left[\sqrt{1-|u_{j}|^{2}} \sqrt{1-|u_{k}|^{2}}+
\frac{1}{2}u_{j}^{*}u_{k} +\frac{1}{2}u_{k}^{*}u_{j} \right],
\label{12}
\end{equation}
where $u_{k} \in \mathbb{C}$, $|u_{k}| \leq 1$, $D_{k}\geq 0$,
$\sum_{k} D_{k} = 1$. Inequality (\ref{12}) follows from
$\mathcal{D} \leq \sum_{k} D_{k} \sqrt{1-|u_{k}|^{2}}$ and
$\mathcal{V}=|\sum_{k} D_{k} u_{k}|$. In \cite{Englert} it is
claimed that because $|u_{k}| \leq 1$, the square brackets in
Eq.(\ref{12}) satisfy $0 \leq \left[\ldots \right] \leq 1$. In
such a case, $\mathcal{D}^2+\mathcal{V}^2\leq \sum_{j,k} D_{j}
D_{k}=\left(\sum_{j} D_{j} \right) \left(\sum_{k} D_{k}
\right)=1$.

Now, $0 \leq \left[\ldots \right] \leq 1$ does not follow from
$|u_{k}| \leq 1$. Indeed, choosing, e.g., $u_{j}=-u_{k}=1$ we get
$\left[\ldots \right] = -1$. Thus, the lower bound in $0 \leq
\left[\ldots \right] \leq 1$ does not hold. Even though this lower
bound is in fact unnecessary for proving the duality relation, the
upper bound, $\left[\ldots \right] \leq 1$, is not obvious and
should be demonstrated.

Alternatively, we can proceed as follows: $\mathcal{V}=|\sum_{k}
D_{k} u_{k}|$ implies $\mathcal{V} \leq \sum_{k} D_{k} |u_{k}|$,
so that $\mathcal{V}^{2}\leq \sum_{j,k} D_{j} D_{k}
|u_{j}||u_{k}|$. Hence,
\begin{equation}
\mathcal{D}^2+\mathcal{V}^2\leq \sum_{j,k}D_{j} D_{k}
\left[\sqrt{1-|u_{j}|^{2}} \sqrt{1-|u_{k}|^{2}}+|u_{j}||u_{k}|
\right]. \label{13}
\end{equation}
The square brackets in (\ref{13}) do satisfy $0 \leq \left[\ldots
\right] \leq 1$. Indeed, the lower bound is obvious, and the upper
bound follows from the Schwarz inequality, which reads $\left(
a_{1}b_{1}+ \ldots + a_{n}b_{n} \right)^{2} \leq \left( a_{1}^{2}+
\ldots + a_{n}^{2} \right) \left( b_{1}^{2}+ \ldots + b_{n}^{2}
\right)$ for reals $a_{i}, b_{i}$. Using this inequality we obtain
\begin{eqnarray}
\left[ \sqrt{1-|u_{j}|^{2}} \sqrt{1-|u_{k}|^{2}} +|u_{j}||u_{k}|
\right]^{2} \leq    \left[ \left( \sqrt{1-|u_{j}|^{2}}
\right)^{2}+|u_{j}|^{2} \right] \left[ \left( \sqrt{1-|u_{k}|^{2}}
\right)^{2}+|u_{k}|^{2} \right]=1,
\end{eqnarray}
so that we can conclude that $\left[\ldots \right] \leq 1$ and the
duality relation is thereby proved.

\end{document}